\documentclass[a4paper,11pt]{article}
\usepackage{jinstpub}
\usepackage{lineno}
\usepackage{multirow}
\usepackage{graphicx}

\title{\boldmath The high speed analog optical readout system designed for low temperature experiments}

\author{Z. Zhou,}
\author[1]{W. Wu,\note{Corresponding author.}}
\author[1]{J. Tang,}
\author{Y. Fu,}
\author{Y. Guo,}
\author{Y. Liu,}
\author{X. Wang,}
\author{W. Zhi}

\affiliation{State Key Laboratory of Dark Matter Physics, Key Laboratory for Particle Astrophysics and Cosmology (MoE), Shanghai Key Laboratory for Particle Physics and Cosmology, School of Physics and Astronomy, Shanghai Jiao Tong University, Shanghai 200240, China}

\emailAdd{wuweihao@sjtu.edu.cn, tangjn@sjtu.edu.cn}

\abstract{
For low-temperature experiments such as liquid xenon dark matter detectors, it is crucial to read out detector signals from cryostats. Traditionally, photoelectrical signals are transmitted from the cryogenic region to the outside using coaxial cables through vacuum feedthroughs on the cryostats. In this paper, we investigate an analog optical transmission method in which the raw electrical signals are converted into optical signals with light intensity linearly proportional to the electrical amplitude, transmitted out of the cryogenic environment through optical fiber, and subsequently converted back into electrical signals by photoelectric devices while preserving the signal waveform. This new approach offers advantages, including low attenuation over long-distance transmission and reduced crosstalk across the feedthroughs. Additionally, the low-temperature optical wavelength multiplexing scheme has been investigated and applied, increasing the transmission capability of a single fiber. At -100~$^{\circ}$C, the proposed analog optical readout system achieves a -3~dB bandwidth of larger than 150~MHz, a dynamic range of up to 500~mV, and a low cryogenic-region power consumption of 70~mW per channel, demonstrating its strong potential for low-temperature experiments.
}


\begin{document}
\maketitle
\flushbottom

\section{Introduction}
\label{sec:intro}

Liquid xenon detectors are a class of cryogenic detectors that are widely used in particle physics experiments, such as the detection of weakly interacting massive particles (WIMPs)~\cite{wimp} and the search for neutrinoless double beta decay (NLDBD)~\cite{NLDBD}. These detectors operate at low temperatures, typically down to about -$100~^{\circ}$C. Therefore, a Dewar structure is employed for thermal insulation, in which the lower volume is filled with liquid xenon (LXe) while the upper volume contains gaseous xenon. Incoming particles deposit their energy in the LXe region. First, prompt scintillation photons are produced in the LXe at a wavelength of 178~nm~\cite{178nmlight1,178nmlight2}. Subsequently, delayed electroluminescence photons are generated in the gaseous region from ionized electrons that drift to the liquid surface and produce a secondary scintillation signal through the electroluminescence process~\cite{S2scintillation}. Both types of photons are detected by photomultiplier tubes (PMTs) or silicon photomultiplier tubes (SiPMs), generating corresponding electrical signals. Currently operating LXe detectors, such as XENONnT~\cite{XENONnT}, LUX-ZEPLIN~\cite{LZ1,LZ2}, and PandaX-4T~\cite{PandaX4T}, which feature hundreds of signal channels, rely on coaxial cables and matched vacuum feedthrough to transmit the raw signals to the outside.

This coaxial-cable readout method directly transmits the raw signals and features simplicity and good robustness. However, in the case of long-distance transmission (tens to hundreds of meters), this method may be bothered by signal attenuation. And crosstalk may be introduced while the signals travel through the vacuum feedthroughs as the central conductor and the grounded outer layer of the coaxial cables are separated and connected to different channels, which degrades the signal integrity. One of the solutions is to use the digital optical readout schemes, which digitize the raw signals at low temperature and convert the electrical signals into optical signals, and it can effectively eliminate attenuation and crosstalk issues. Nevertheless, such approaches introduce additional system complexity, lowering the robustness. The low-temperature digitization also increases the power consumption in the cryogenic region, which may impose significant challenges on the detector refrigeration system. To address these limitations, we have developed a low-temperature analog optical readout method that significantly reduces long-distance attenuation and feedthrough crosstalk while consuming substantially less power and bringing more simplicity than digital solutions, thereby providing an alternative readout option for LXe experiments.

\begin{figure}[htbp]
    \centering
    \includegraphics[width=0.9\linewidth]{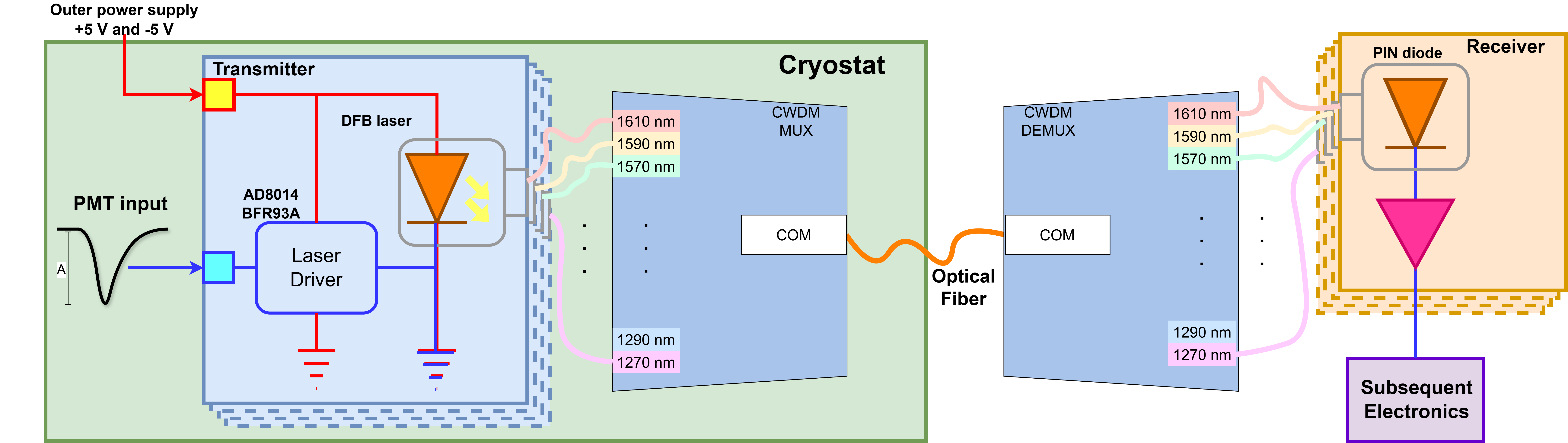}
    \caption{The multiplexed analog optical readout diagrams.}
    \label{fig:diagram}
\end{figure}

In the new readout approach, the input electrical signals are converted into optical signals by customized-designed optical transmitters, in which the real-time optical power is linearly related to the amplitude of the input signals. The converted signals are then transmitted through optical fibers to the room temperature environment. Outside the cryostat, optical receivers convert the optical signals back into electrical signals, which are subsequently sent to subsequent electronics. Similar approaches have been explored by the DUNE~\cite{DUNE1} and DarkSide-20k~\cite{DarkSide20k} collaborations. Building upon these efforts, our work introduces a simplified transmitter architecture with optimized circuit parameters, enabling a significantly higher analog bandwidth and lower power dissipation. In addition, the application of optical wavelength division multiplexing (WDM) at low temperature is studied and experimentally implemented, allowing multiple optical signals at distinct wavelengths to be transmitted through a single optical fiber and thereby improving the transmission capacity of a single fiber. A schematic diagram of the multiplexed analog optical readout method operation is shown in Fig.~\ref{fig:diagram}.

In this article, we explore the new approach of a low-temperature analog optical readout method and perform the performance characterization of the prototype. The performance requirements of the readout system for LXe experiments (as an example) are discussed in Section~\ref{sec:requirements}. The design details of the optical transmitters and receivers, as well as their performance evaluation, are described in Section~\ref{sec:design-scheme}. The application of low-temperature multiplexing and the PMT-prototype commissioning are reported in Section~\ref{sec:WDM}, and the conclusions and discussions are given in Section~\ref{sec:sum}.

\section{Performance requirements}
\label{sec:requirements}

In this section, we discuss the performance requirements of the analog optical readout system for LXe dark matter experiments. The primary electrical characteristics of concern include the -3~dB bandwidth, the dynamic range, and the single photo-electron (SPE) signal-to-noise ratio (SNR). In addition, the low-temperature operation of LXe detectors imposes stringent constraints on the power consumption of the cryogenic electronics. Overall, the analog optical readout system should satisfy the following requirements, which are summarized in Tab.~\ref{tab:requirements}:

\begin{enumerate}

    \item The -3~dB bandwidth of the readout system is a critical parameter that determines the timing precision of the signals, while a higher bandwidth also improves the capability of accurately reproducing signal waveforms. However, increasing the bandwidth generally leads to higher power consumption and lower gain, which can adversely affect low-temperature operation and the SNR. Therefore, an appropriate balance must be achieved. For the PMTs commonly used in LXe experiments, such as the Hamamatsu R11410~\cite{R11410}, a bandwidth larger than 100~MHz is considered suitable.

    \item The dynamic range determines the maximum signal amplitude that the system can process, and a larger dynamic range is beneficial for detecting high-energy events in LXe detectors. However, the dynamic range cannot be increased arbitrarily due to the maximum input amplitude ratings of subsequent electronics, such as amplifiers and ADCs, and by its inherent trade-off with the SNR. Considering these factors, an operationally reasonable dynamic range for the optical readout system is about 500~mV, corresponding to approximately 80 times the SPE amplitude. The SNR is therefore treated as a coupled design parameter rather than a strict requirement, and an average SNR exceeding 4 is considered sufficient for reliable SPE identification.
    
    \item Since the system is designed for low-temperature operation, its power consumption must be carefully controlled to reduce the cooling load of the LXe detector and to avoid the risk of local heating that could induce bubble formation in the LXe time projection chamber (TPC). As a reference, the power consumption of the readout system should be comparable to that of the PMT base boards~\cite{HuangDi}, which are also operated at cryogenic temperatures, and should be kept below 100~mW per channel.
\end{enumerate}

\begin{table}[htbp]
    \caption{The requirements of the analog optical readout system for the LXe experiments.}
    \label{tab:requirements}
    \centering
    \resizebox{\textwidth}{!}{
    \begin{tabular}{c|cccc}
    \hline
    Characteristics    & -3~dB bandwidth & Dynamic range & SNR     & Power consumption \\
    \hline
    Requirements       & $\ge$100~MHz  & $\ge$500~mV & $\ge$4  & $\le$100~mW per channel\\
    \hline
    \end{tabular}
    }
\end{table}

\section{Design scheme and evaluation}
\label{sec:design-scheme}

The low temperature analog optical readout system can be divided into the analog optical transmitter that operates at low temperature, and the analog optical receiver that operates in room temperature.

\subsection{The analog optical transmitter design}
\label{subsec:transmitter}

\begin{figure}[htbp]
    \centering
    \includegraphics[width=0.48\linewidth]{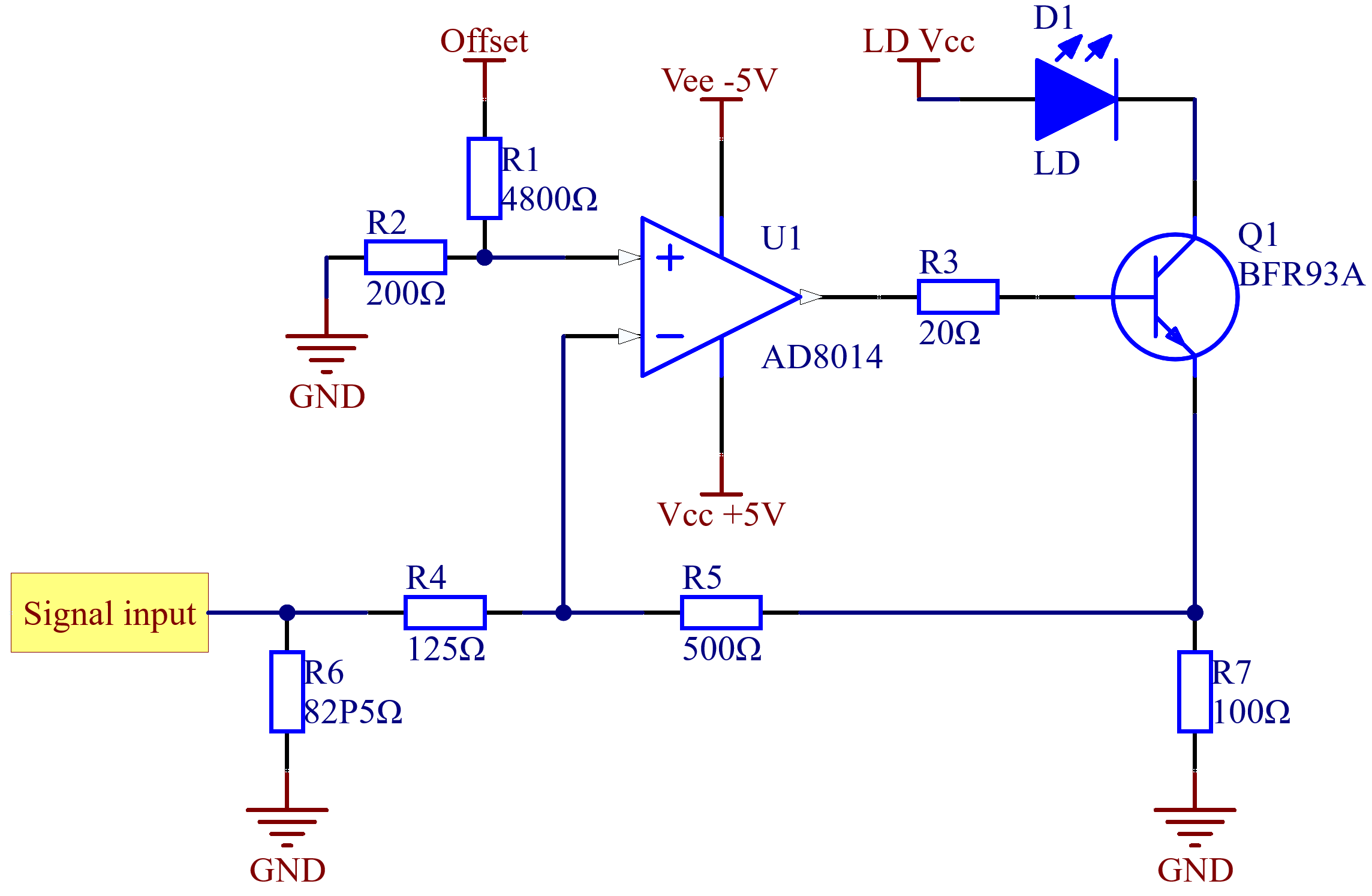}
    \quad
    \includegraphics[width=0.48\linewidth]{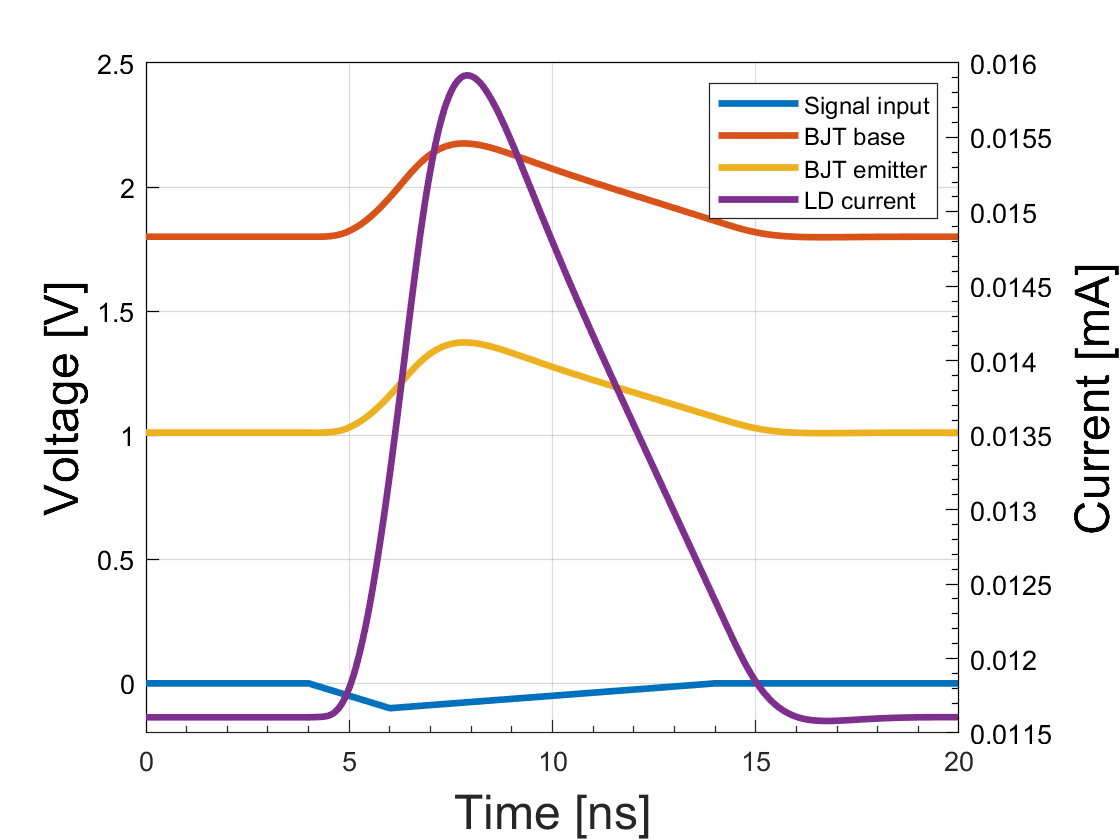}
    \caption{(Left) The schematic diagram of one of the channels of the optical transmitter. The laser diodes are LSCDLDX-4-S-0-2-fJKFC, produced by Beijing Lightsensing Technologies Ltd, where X is the corresponding wavelength. (Right) The Spice simulation result of a 100~mV SPE-like signal input, with a simulation time step of 0.01~ns.}
    \label{fig:sch-transmitter}
\end{figure}

The analog optical transmitter is responsible for converting the input electrical signals from the photoelectric detectors into optical signals, with the emitted optical power linearly proportional to the amplitude of the electrical input. The schematic diagram of the transmitter is shown in Fig.~\ref{fig:sch-transmitter} (Left).

The electrical signal from the detector is first inverted and amplified by an operational amplifier (op-amp). The closed-loop gain of the op-amp is set to 4, representing an optimized trade-off between the SNR and the dynamic range, while maintaining sufficient phase margin as the operating temperature decreases. A wide-band bipolar junction transistor (BJT), with its base and emitter connected within the feedback loop of the op-amp and its collector connected to the cathode of the laser diode (LD), is employed to linearly convert the amplified voltage signal into the driving current of the laser diode, as shown in the Spice simulation result shown in Fig.~\ref{fig:sch-transmitter} (Right). The base-emitter turn-on voltage $V_{BE(\mathrm{on})}$ of the BJT is selected to be as low as possible to maximize the usable output voltage swing of the op-amp and thereby increase the dynamic range.

At the non-inverting input of the op-amp, a voltage divider is implemented to provide a DC offset, which is converted into the threshold current of the laser diode. This biasing scheme ensures that the laser diode operates above threshold and enables an immediate optical response to the input electrical signals.

The power consumption of the transmitter consists of contributions from the op-amp quiescent power and the laser diode threshold power. While the former can only be reduced through the selection of suitable low-power op-amps, the latter can be minimized by employing LDs with low threshold currents and by reducing the bias voltage applied to the laser diode anode. With these measures, the power consumption of the analog optical transmitter is reduced to 65~mW per channel at room temperature, and increases slightly to 70~mW per channel at -100~$^{\circ}$C.

With respect to the dynamic range, the limiting factors are more complex. As the amplitude of the input signal $V_{in}$ increases, the transmitter may enter saturation or deviate from linear operation under the following conditions:
\begin{enumerate}
    \item As $V_{in}$ increases, the voltage at the BJT emitter rises accordingly, leading to a reduction of the collector-emitter voltage $U_{ce}$. Consequently, the operating region of the BJT may transition from the active region to saturation. This condition can be avoided by increasing the bias voltage applied to the laser diode anode, thereby maintaining the BJT in the active operating region.

    \item The collector-emitter current $I_{ce}$, which is approximately equal to the laser diode current $I_{LD}$, may exceed the maximum allowable current of the laser diode as $V_{in}$ increases. This limitation can be mitigated by increasing the value of the load resistor R7 connected between the BJT emitter and ground.

    \item The base-emitter voltage $U_{be}$ increases with $V_{in}$, and the corresponding base voltage of the BJT may eventually reach the output swing limit of the op-amp, resulting in saturation of the transmitter.
\end{enumerate}
To maximize the achievable dynamic range of the transmitter while maintaining low power consumption, saturation conditions 1 and 2 are avoided through optimization of the bias voltage applied to the laser diode anode and the value of the emitter load resistor R7. Saturation condition 3 can be further alleviated by employing an op-amp with a wider output voltage swing. However, this approach generally leads to increased power consumption and is therefore not adopted in the present design.

\subsection{The analog optical receiver design}
\label{subsec:receiver}

\begin{figure}[htbp]
    \centering
    \includegraphics[width=0.9\linewidth]{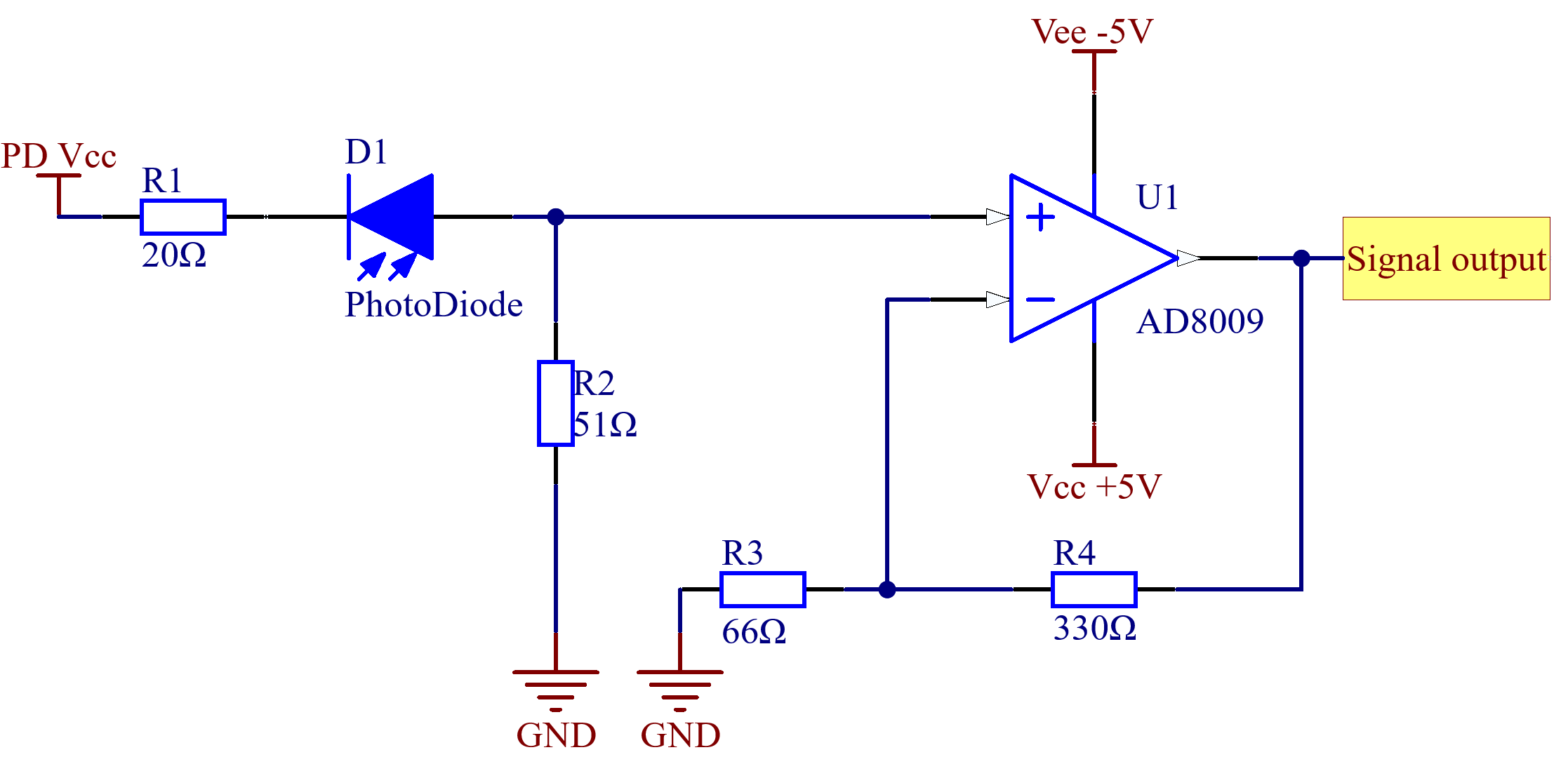}
    \caption{The schematic diagram of one of the channels of the optical receiver. The photodiodes are the LSIPD-A75-B-2JKFC produced by Beijing Lightsensing Technologies Ltd.}
    \label{fig:sch-receiver}
\end{figure}

The analog optical receiver is designed to convert the optical signals transmitted from the cryostat back into electrical signals at room temperature. Since there are no power consumption constraints outside the cryogenic environment, high-speed components with a wide dynamic range can be employed in the receiver design without limitation.

The receiver consists of a high-speed Si PIN photodiode and a corresponding amplifier circuit, as shown in Fig.~\ref{fig:sch-receiver}. The photocurrent generated by the photodiode is directly passed through a 51~$\Omega$ resistor, converting it into a voltage signal. This voltage signal is subsequently amplified by a factor of 6 and delivered to the downstream electronics. By appropriate selection of the photodiode and by applying a sufficiently high reverse bias voltage, the receiver avoids saturation over the full dynamic range of the transmitter.

\subsection{Performance evaluation}
\label{subsec:evaluate}

\begin{figure}[htbp]
    \centering
    \includegraphics[width=0.5\textwidth]{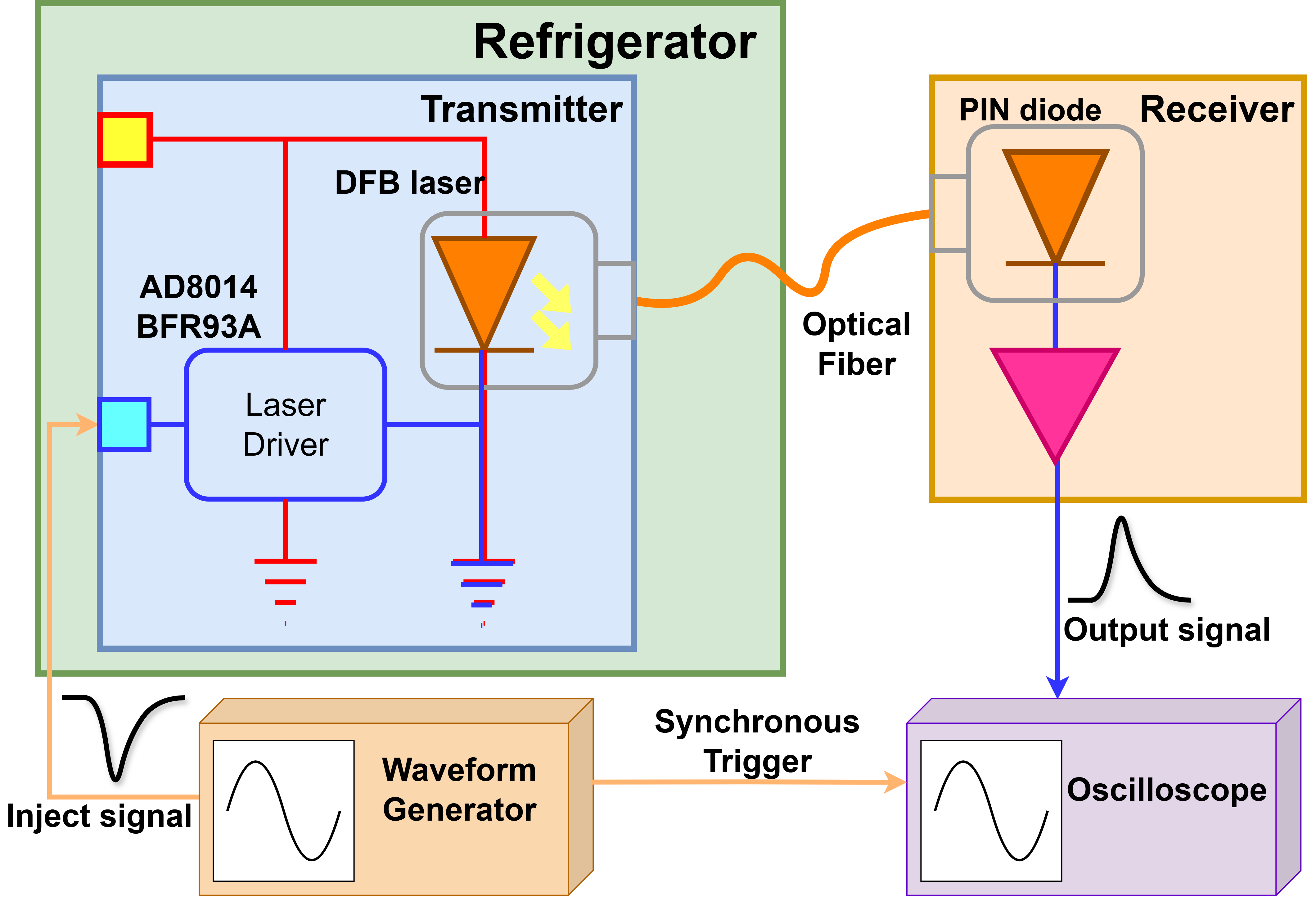}
    \quad
    \includegraphics[width=0.46\textwidth]{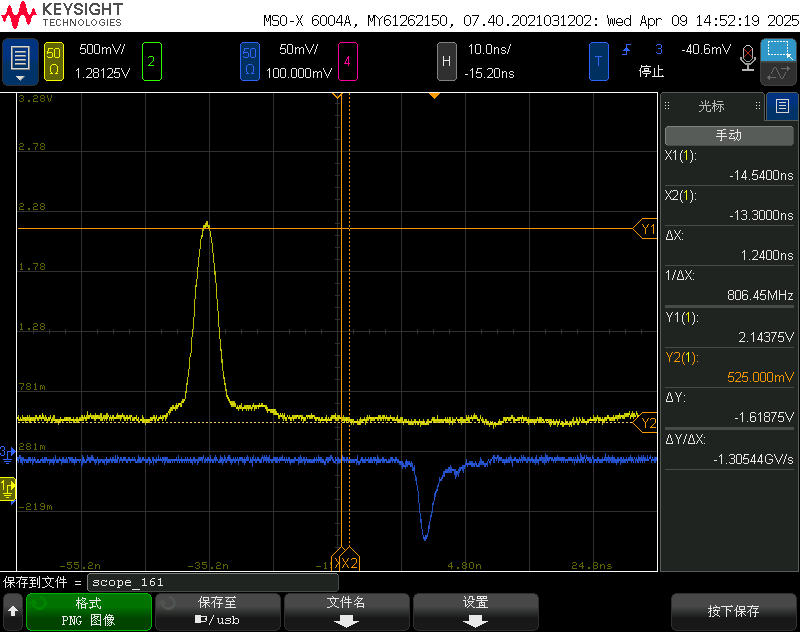}
    \caption{(Left) The test diagram of the analog optical readout system performance evaluation. (Right) The receiver output (yellow) as a SPE-like waveform (blue) is injected by a waveform generator.}
    \label{fig:system_eval}
\end{figure}

The performance evaluation of the analog optical readout system with a Tektronix AFG31252 waveform generator signal input was carried out over a temperature range from room temperature down to -120~$^{\circ}$C, and the characteristics at different temperatures were systematically compared. In the refrigerator, additional PT100 temperature sensors were installed to provide more accurate temperature monitoring. During the tests, the transmitter of the analog optical readout system was placed inside the cryogenic volume of the refrigerator, while the receiver remained outside at room temperature. The optical transmitter and receiver were connected by a single-mode (SM) optical fiber, which passed through a dedicated feedthrough hole on the side wall of the refrigerator. The diagram of the evaluation is shown in Fig.~\ref{fig:system_eval} (Left), and the response to the SPE-like signal input, shown in Fig.~\ref{fig:system_eval} (Right), verified the functionality of this prototype.

\subsubsection{Bandwidth}
\label{subsubsec:BW}

\begin{figure}[htbp]
    \centering
    \includegraphics[width=0.6\textwidth]{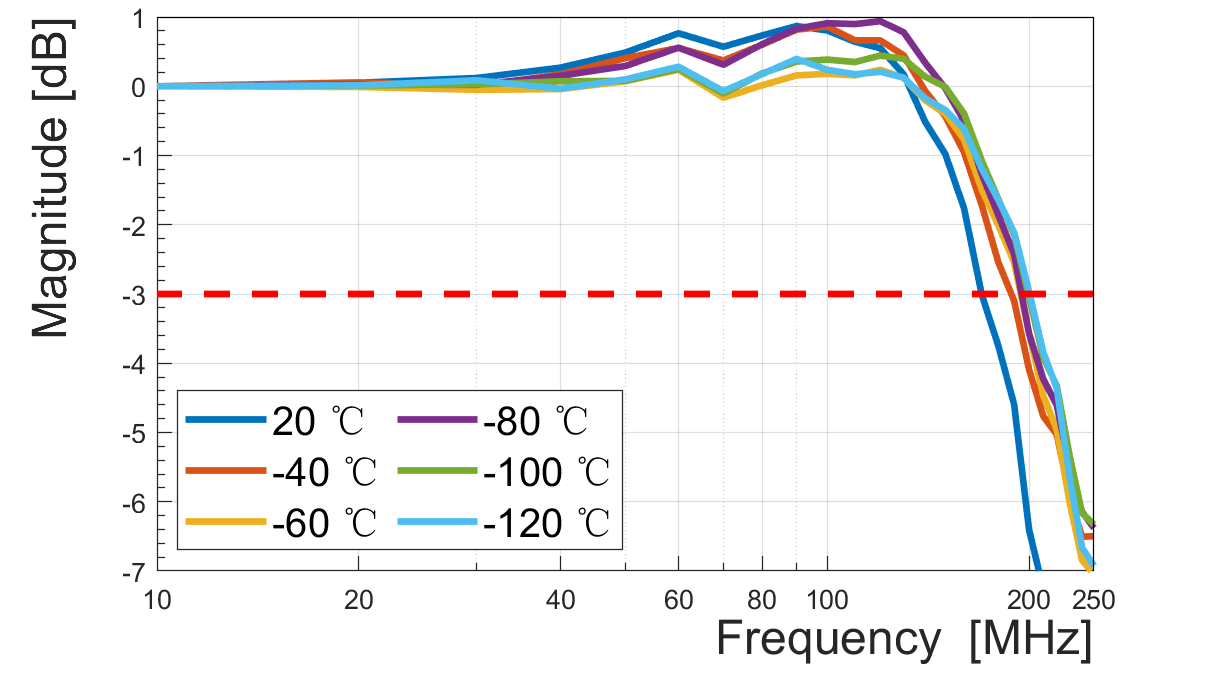}
    \caption{The Bode plot of the analog optical readout system, and the -3~dB frequency is around 170~MHz at room temperature and increases to about 190~MHz at -40~$^{\circ}$C, -60~$^{\circ}$C, and -80~$^{\circ}$C, and 200~MHz at -100~$^{\circ}$C and -120~$^{\circ}$C.}
    \label{fig:Bode}
\end{figure}

The -3~dB bandwidth of the analog optical readout system is measured by injecting continuous sine signals with frequencies from 10~MHz to 250~MHz (the fastest signal that our waveform generator is able to produce) into the transmitter and recording the corresponding peak-to-peak value at the receiver output using a Keysight MSOX6004A 2.5 GHz oscilloscope (20 GS/s), and the transmitter and receiver are connected through an SM optical fiber. At different temperatures, the Bode plot of the system is shown in Fig.~\ref{fig:Bode}. At room temperature, the system exhibits a bandwidth of 170~MHz. As the temperature gradually goes down to -120~$^{\circ}$C, the bandwidth of the system increases to 200~MHz at most.

\subsubsection{Dynamic range}
\label{subsubsec:DR}

\begin{figure}[htbp]
    \centering
    \includegraphics[width=0.48\textwidth]{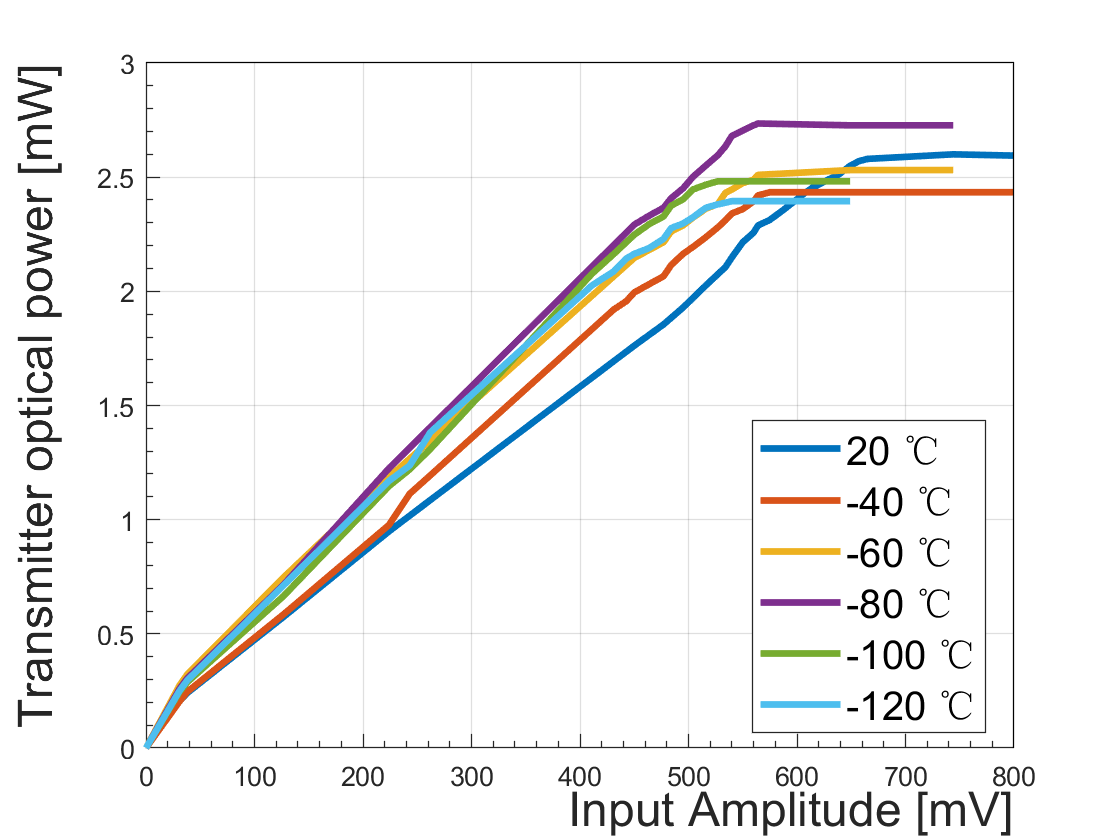}
    \quad
    \includegraphics[width=0.48\textwidth]{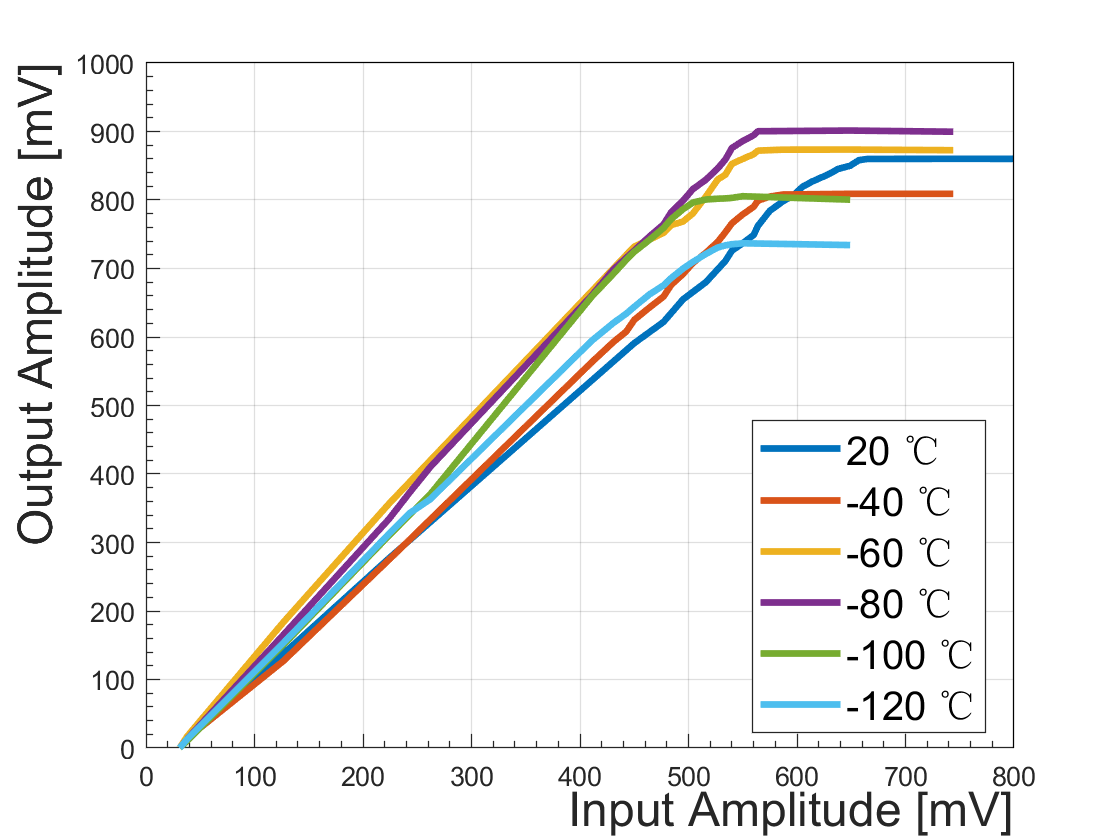}
    \caption{(Left) The measured optical power of the optical transmitter output. (Right) The measured signal amplitude of the optical receiver output.}
    \label{fig:amp_DR}
\end{figure}

The dynamic range of the analog optical readout system is also measured in the condition of connecting the transmitter and receiver with an SM optical fiber. During the measurement, direct-current (DC) are injected into the transmitter, and an oscilloscope is responsible for reading the output amplitudes from the receiver. The output optical powers from the transmitter are also measured for further verification. The excess of the dynamic range is determined as the output amplitudes deviate from the linear relation with the input signals. The results indicate that the output optical power of the optical transmitter begins to deviate from linearity at an input amplitude of approximately 650~mV at room temperature, as shown in Fig.~\ref{fig:amp_DR} (Left), corresponding to about 100 times the amplitude of a single photoelectron (SPE) signal. The dynamic range decreases as the temperature goes down, which is 540~mV at -40~$^{\circ}$C to -80~$^{\circ}$C, and further decreases to 500~mV at -100~$^{\circ}$C and -120~$^{\circ}$C. The measurement result of the whole system, as shown in Fig.~\ref{fig:amp_DR} (Right), consists with the optical power result, showing no saturation at the optical receiver.

\section{Channel multiplexing at low temperature}
\label{sec:WDM}

To increase the signal transmission capacity of the single optical fiber, coarse wavelength division multiplexing (CWDM) modules together with narrow-spectral-linewidth distributed feedback (DFB) laser diodes were investigated for operation at low temperature, and four-channel-multiplexing was successfully achieved at -100~$^{\circ}$C. The main challenge of low-temperature multiplexing is the potential mismatch between the emission wavelengths of the DFB lasers and the corresponding CWDM channels, since the laser wavelength shifts as the temperature decreases~\cite{coldLDwavelength}. Therefore, evaluating the behavior of both the CWDM modules and the DFB lasers down to -100~$^{\circ}$C is essential.

\begin{figure}[tbp]
    \centering
    \includegraphics[width=0.9\textwidth]{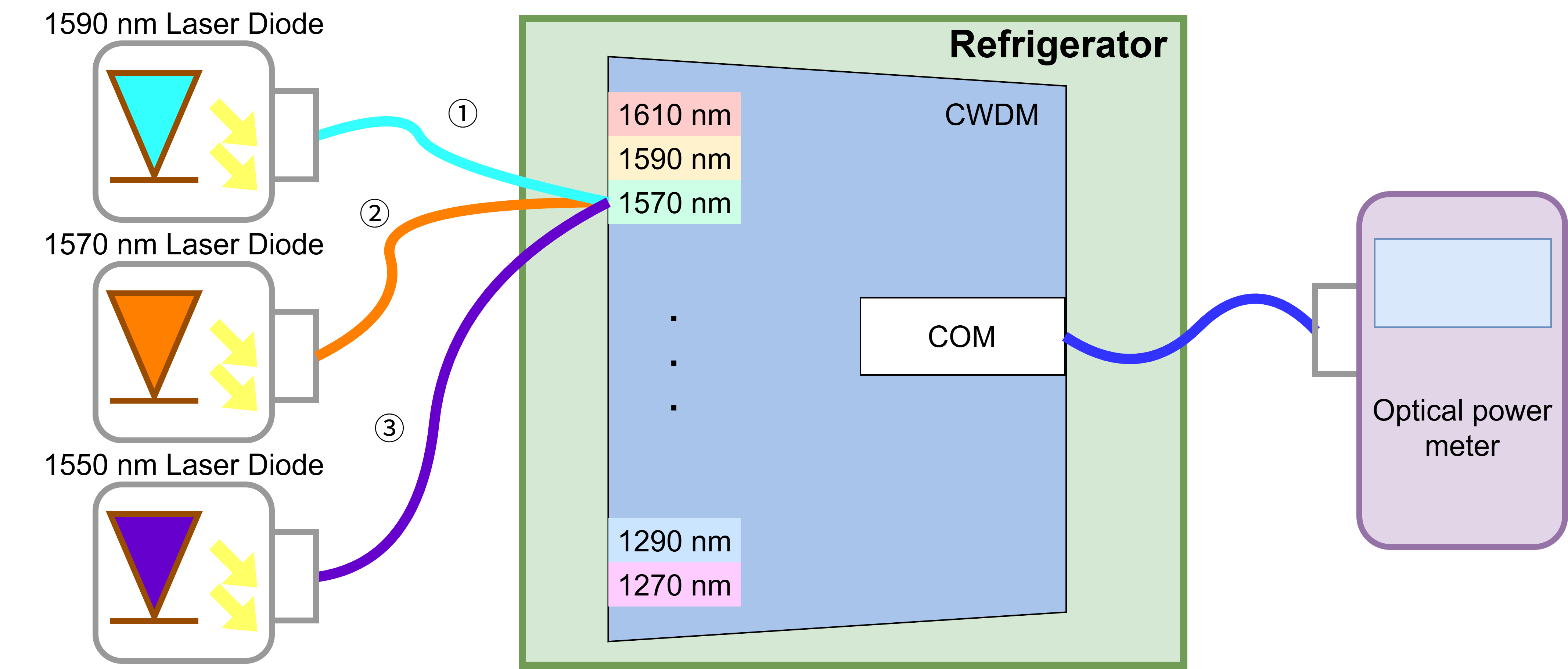}
    \caption{Test diagram of the CWDM low-temperature test. To evaluate the effect of cooling on the CWDM channel (such as the 1570~nm channel), we inject light of 1590~nm, 1570~nm, and 1550~nm, respectively, and measure the output optical power at the COM end at different temperatures.}
    \label{fig:CWDMtest}
\end{figure}

\begin{table}[htbp]
        \caption{The measured optical power of COM port at different temperatures as 1550~nm, 1570~nm, and 1590~nm laser lights are input from the 1570~nm channel.} 
        \label{tab:CWDMtest}
        \centering
        \resizebox{\textwidth}{!}{
        \begin{tabular}{ccccccccc}
            \hline
            \multirow{2}{*}{Input wavelengths [nm]} &   \multicolumn{8}{c}{The measured optical power at different temperatures [$\mu$W]}   \\
            \cline{2-9}
                & 20~$^{\circ}$C  & 0~$^{\circ}$C &  -20~$^{\circ}$C & -40~$^{\circ}$C & -60~$^{\circ}$C& -80~$^{\circ}$C & -100~$^{\circ}$C & -120~$^{\circ}$C   \\
            \hline
            1550  & 2.3  & 2.2  & 2.2  & 2.02 & 1.3 & 0.97 & 0.95 & 0.73 \\
            1570  & 643  & 630  & 590  & 589 & 503  & 420  & 363  & 320  \\
            1590  & 0.83 & 0.81 & 0.75 &0.64 & 0.59 & 0.52 & 0.39 & 0.25 \\
            \hline
        \end{tabular}
        }
\end{table}

\subsection{Channel rematching at low temperature}
\label{subsec:rematch}

\begin{figure}[htbp]
    \centering
    \includegraphics[width=0.9\textwidth]{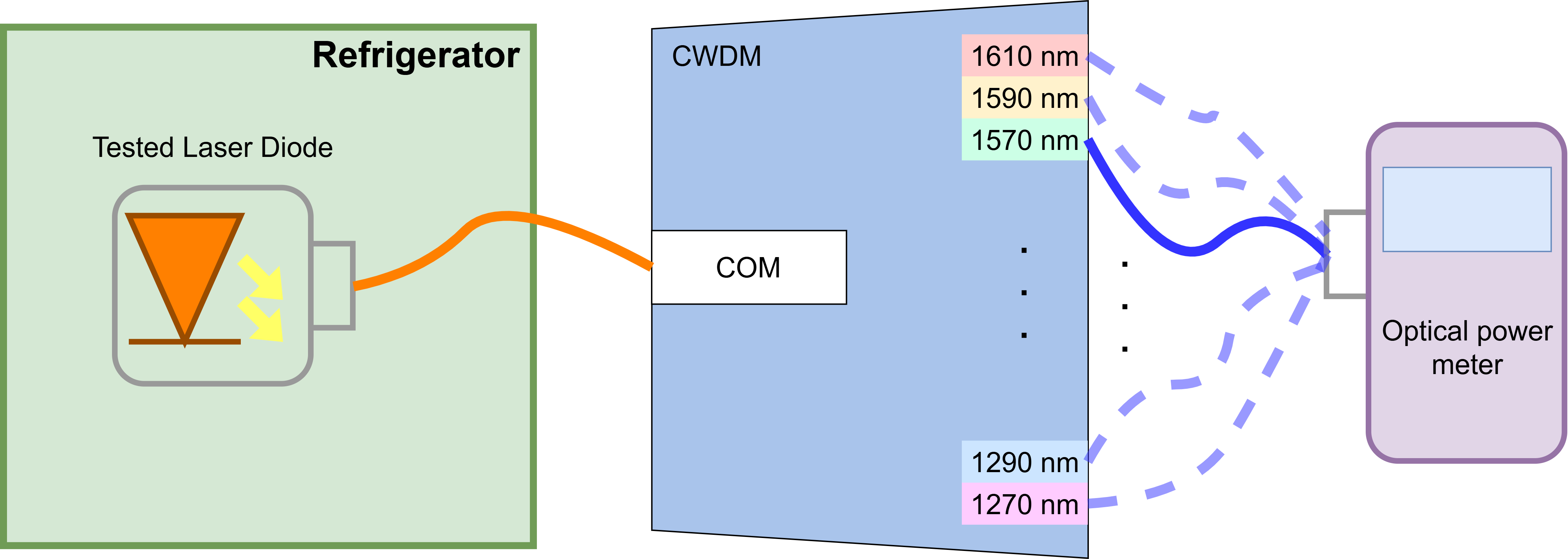}
    \caption{Test diagram of the LDs low temperature test.}
    \label{fig:LDtest}
\end{figure}

In this work, a fiber-cascade-type CWDM, which is the most commonly available configuration, was tested, and its response test to stable laser inputs is shown in Fig.~\ref{fig:CWDMtest}. The measured results for the 1570~nm channel, taken as a representative example, are summarized in Tab.~\ref{tab:CWDMtest}. The results indicate that the passband wavelength of the CWDM remains stable with temperature, and the module continues to function properly at low temperatures. However, the insertion loss increases significantly, such that the transmitted optical power of the nominal wavelength is reduced by 50~\% at low temperature. This effect degrades the SNR in practical applications.

\begin{figure}[htbp]
    \centering
    \includegraphics[width=0.9\textwidth]{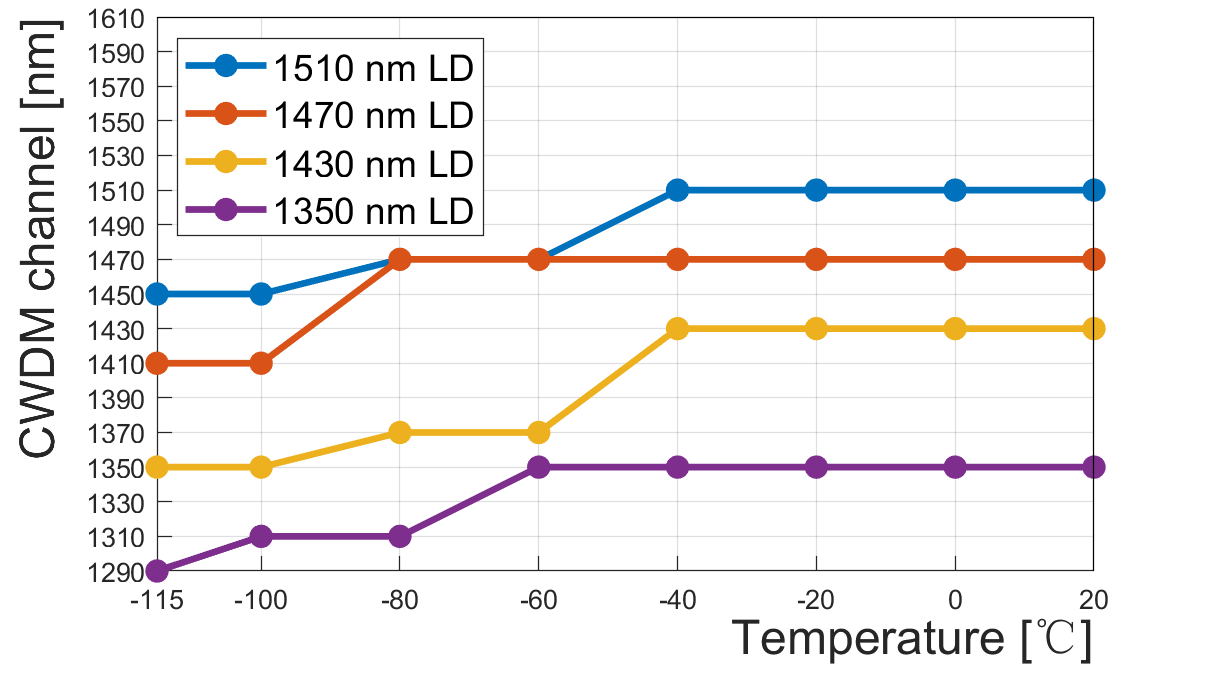}
    \caption{The corresponding CWDM channels where the maximum optical power is measured of the 1510~nm, 1470~nm, 1430~nm and 1350~nm laser diodes as the temperature goes down.}
    \label{fig:LDtest-result}
\end{figure}

To measure the wavelength shift of the light sources at low temperature, laser diodes with nominal wavelengths of 1510~nm, 1470~nm, 1430~nm and 1350~nm were placed inside the refrigerator, and their emitted light was guided into the COM port of a CWDM module located at room temperature. An optical power meter was used to measure the average output optical power of each CWDM channel, as illustrated in Fig.~\ref{fig:LDtest}.

As the temperature decreases, the central wavelengths of the tested laser diodes shift substantially, and the CWDM channels with the maximum optical power, corresponding to the best wavelength matching,  exhibit a blue shift of 40 to 80~nm relative to the nominal laser wavelengths, as shown in Fig.~\ref{fig:LDtest-result}. The results indicate that wavelength-shifted laser diodes at low temperature can be re-matched to the CWDM channels that are blue-shifted relative to the room temperature matching.

\subsection{Commissioning with PMT at low temperature}
\label{subsec:PMT-joint-debug}

\begin{figure}[htbp]
    \centering
    \includegraphics[width=0.46\textwidth]{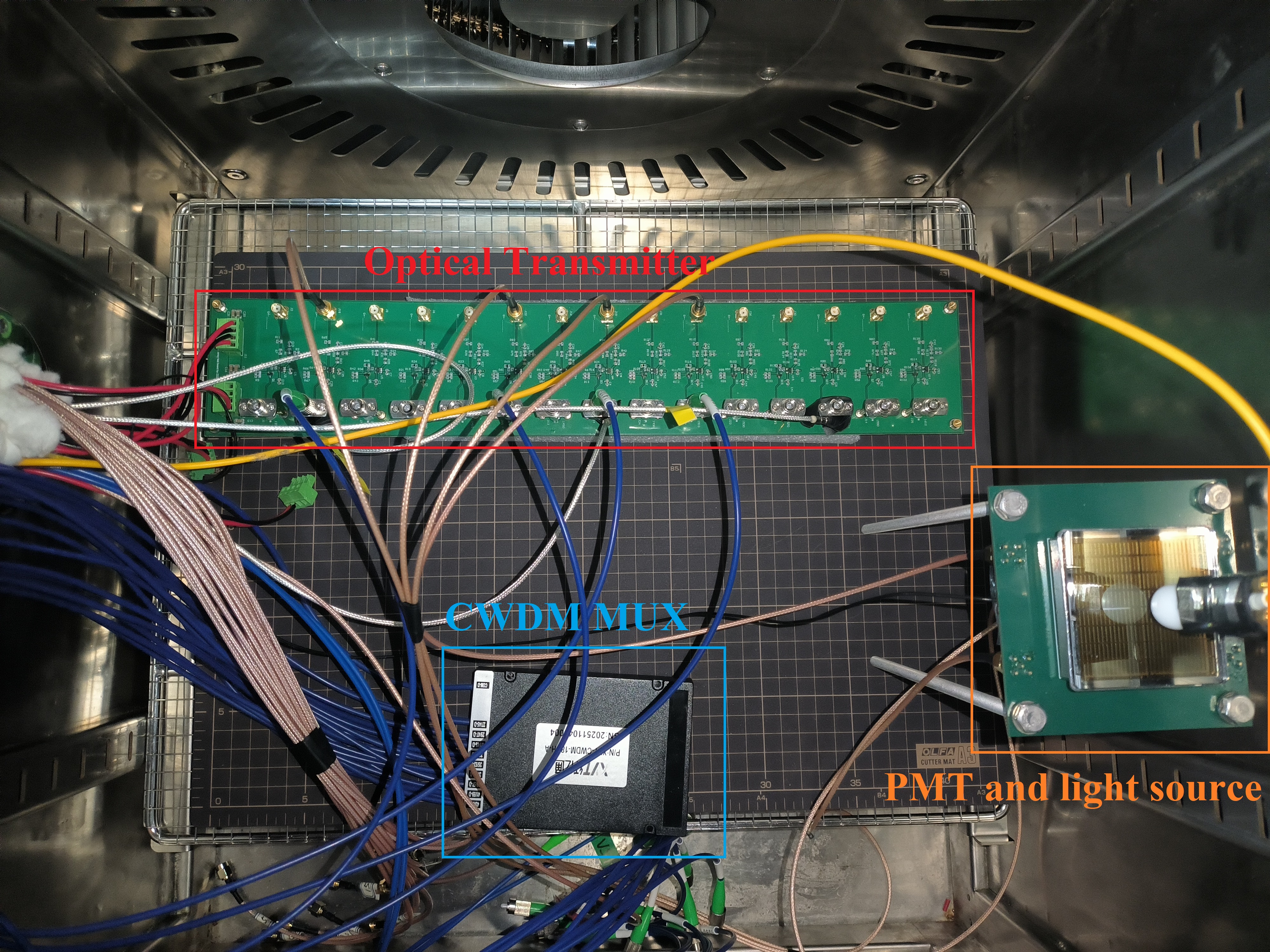}
    \quad
    \includegraphics[width=0.5\textwidth]{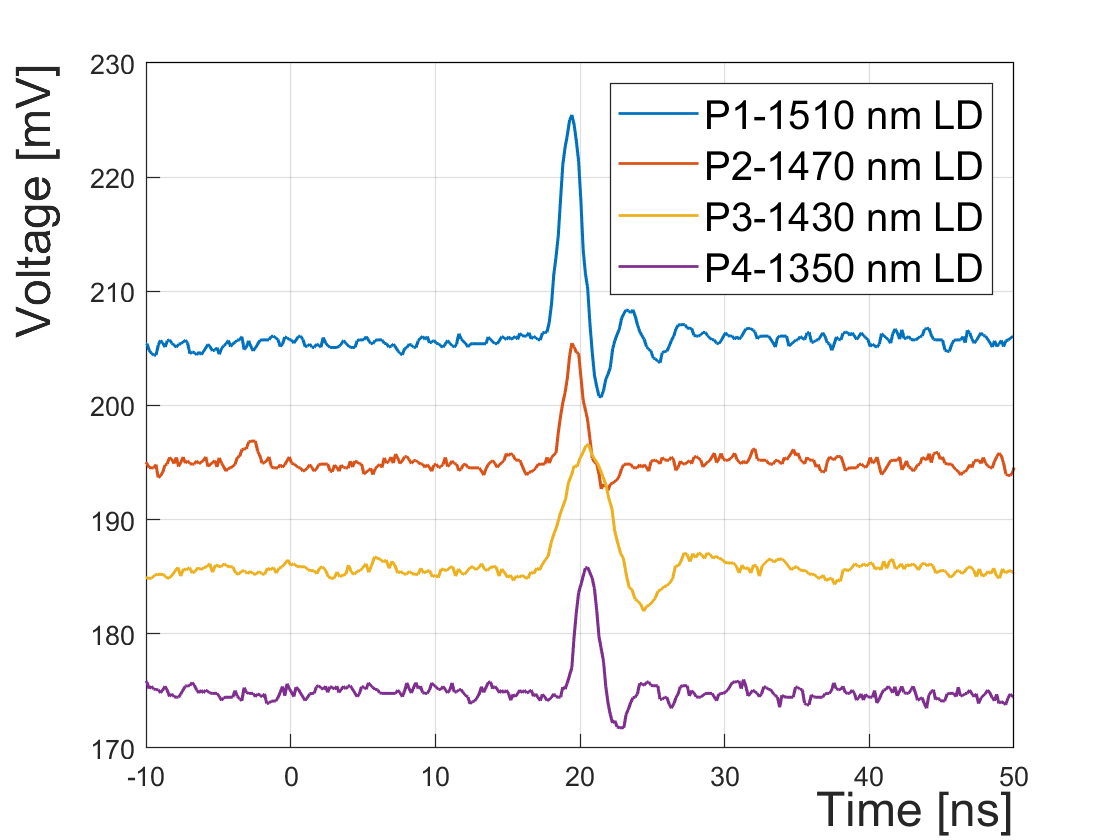}
    \caption{(Left) The cryogenic test platform, the PMT, analog optical readout transmitter, and CWDM MUX are placed inside the refrigerator. (Right) The oscilloscope waveforms (time and baseline are normalized) of the PMT that are read out by the multiplexed analog optical readout prototype.}
    \label{fig:test_platform_waveform}
\end{figure}

The low-temperature pmt-prototype commissioning is deployed close to the real operating condition, the Hamamatsu R12699 PMT signals are injected into the transmitter, while the latter is connected with a CWDM MUX, and all of them are placed into the refrigerator and cooled down to -100~$^{\circ}$C. The COM port of the CWDM MUX penetrates through the opening on the refrigerator, and connects to a CWDM DEMUX, which finally sends the optical signals to the receivers. Fig~.\ref{fig:test_platform_waveform} (Left) shows the setup inside the refrigerator, and the readout PMT signals are shown in Fig~.\ref{fig:test_platform_waveform} (Right). The coaxial cable readouts of the same R12699 PMT are also implemented to compare.

\begin{figure}[tbp]
  \centering
  \vspace{2mm}
  \begin{minipage}{0.49\textwidth}
    \centering
    \includegraphics[width=\linewidth]{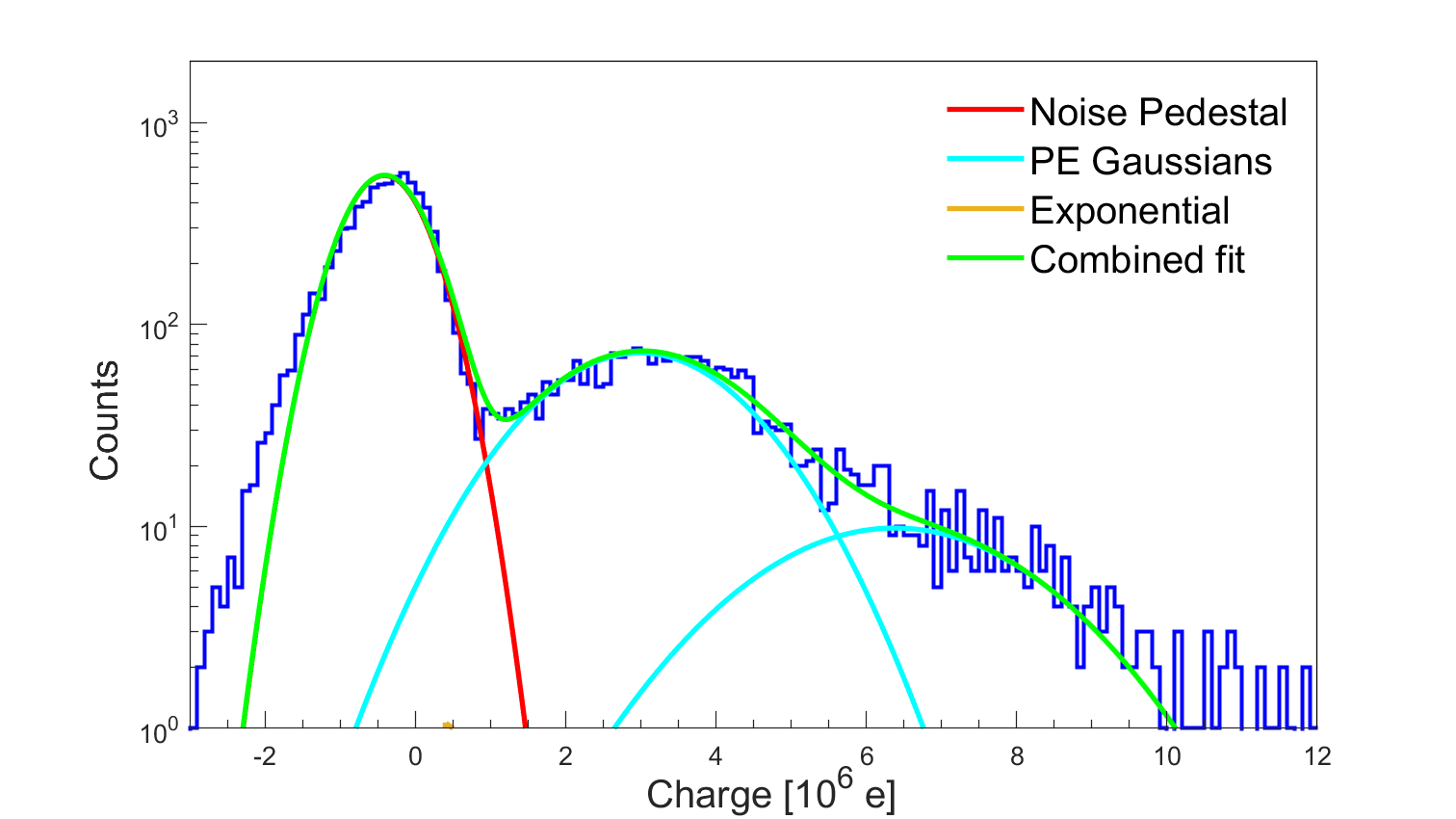}
    {\small\textbf{(a)}}\\[-0.5mm]
  \end{minipage}\hfill
  \begin{minipage}{0.49\textwidth}
    \centering
    \includegraphics[width=\linewidth]{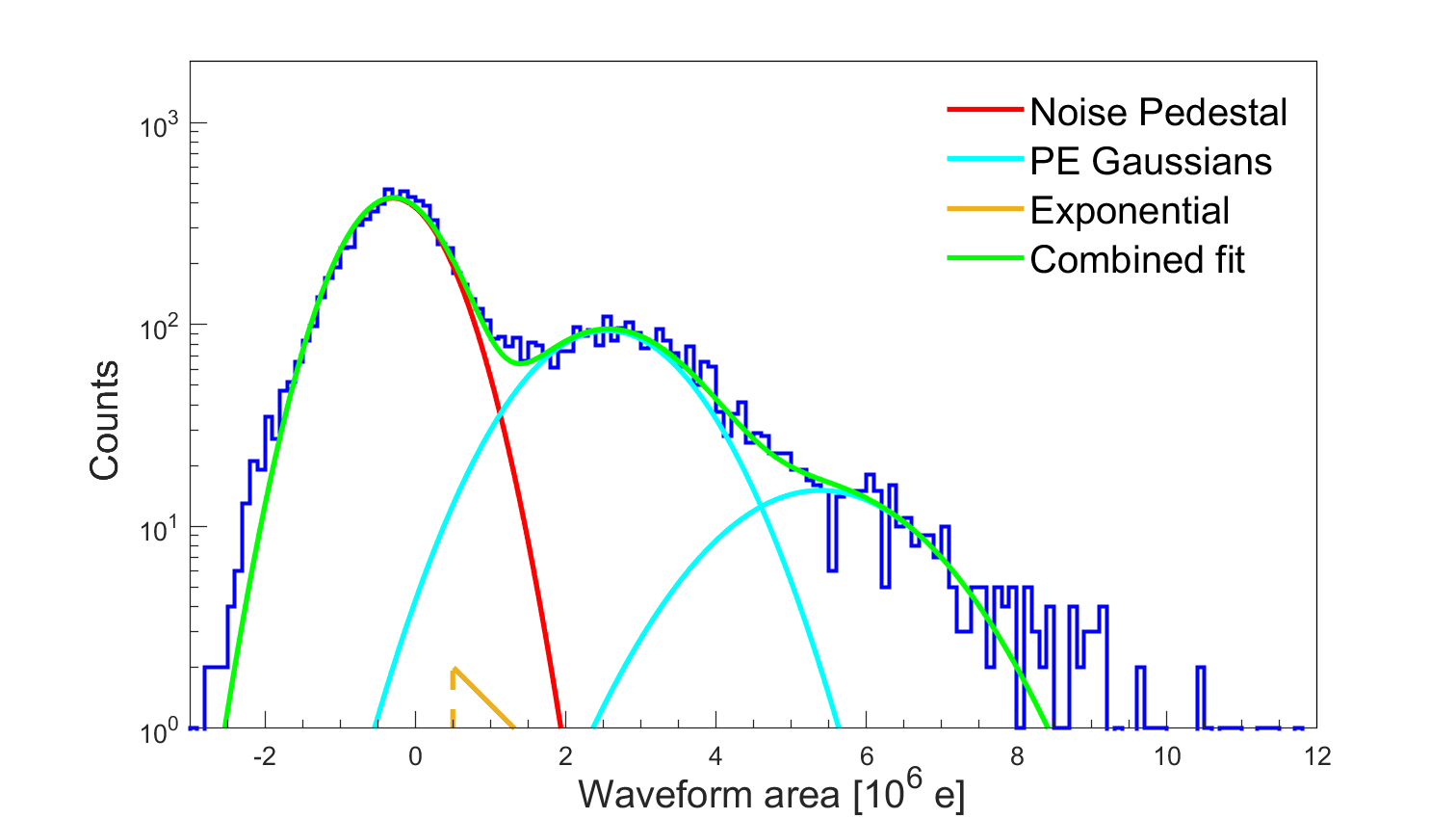}
    {\small\textbf{(e)}}\\[-0.5mm]
  \end{minipage}

  \vspace{2mm}
  \begin{minipage}{0.49\textwidth}
    \centering
    \includegraphics[width=\linewidth]{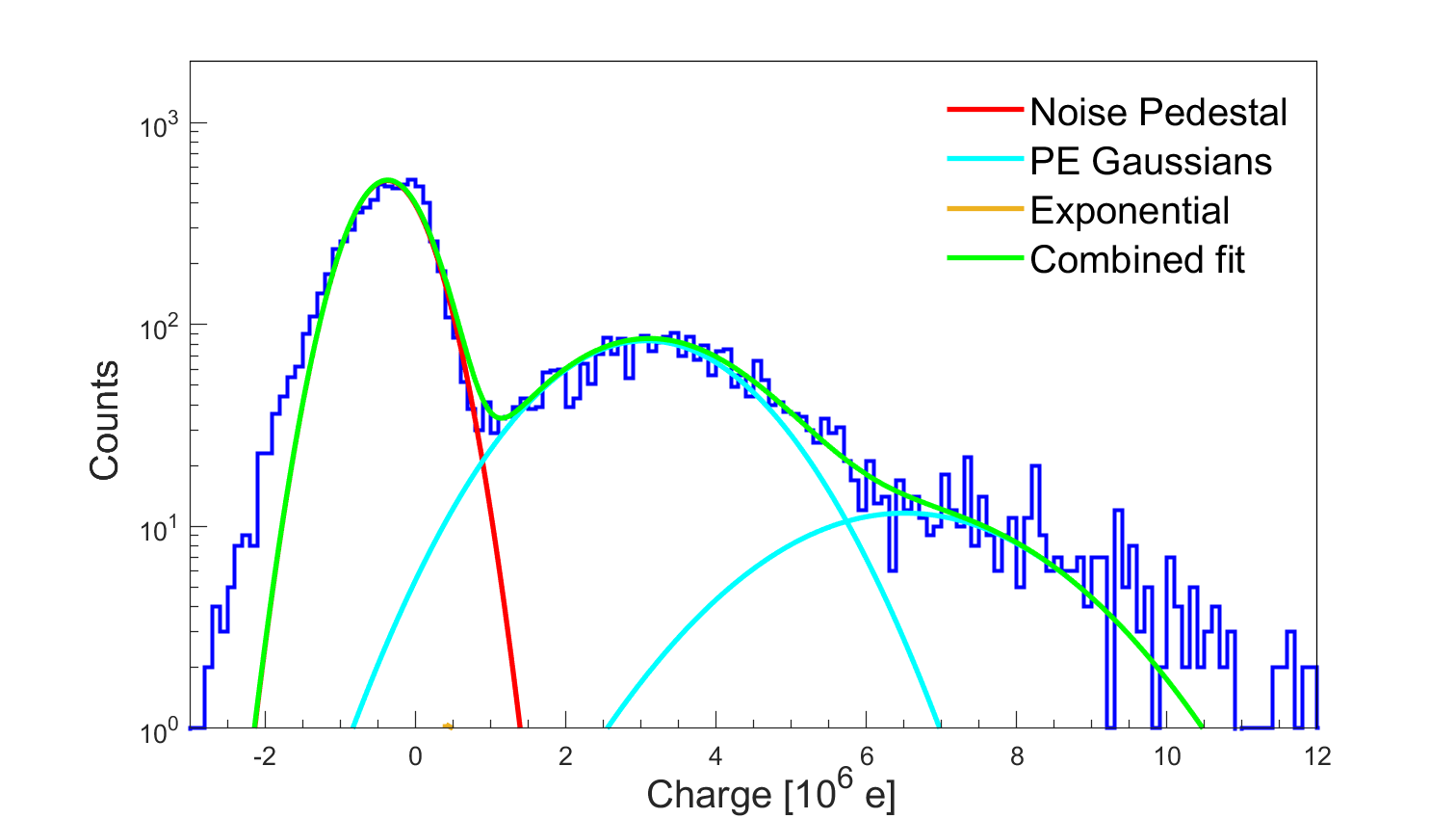}
    {\small\textbf{(b)}}\\[-0.5mm]
  \end{minipage}\hfill
  \begin{minipage}{0.49\textwidth}
    \centering
    \includegraphics[width=\linewidth]{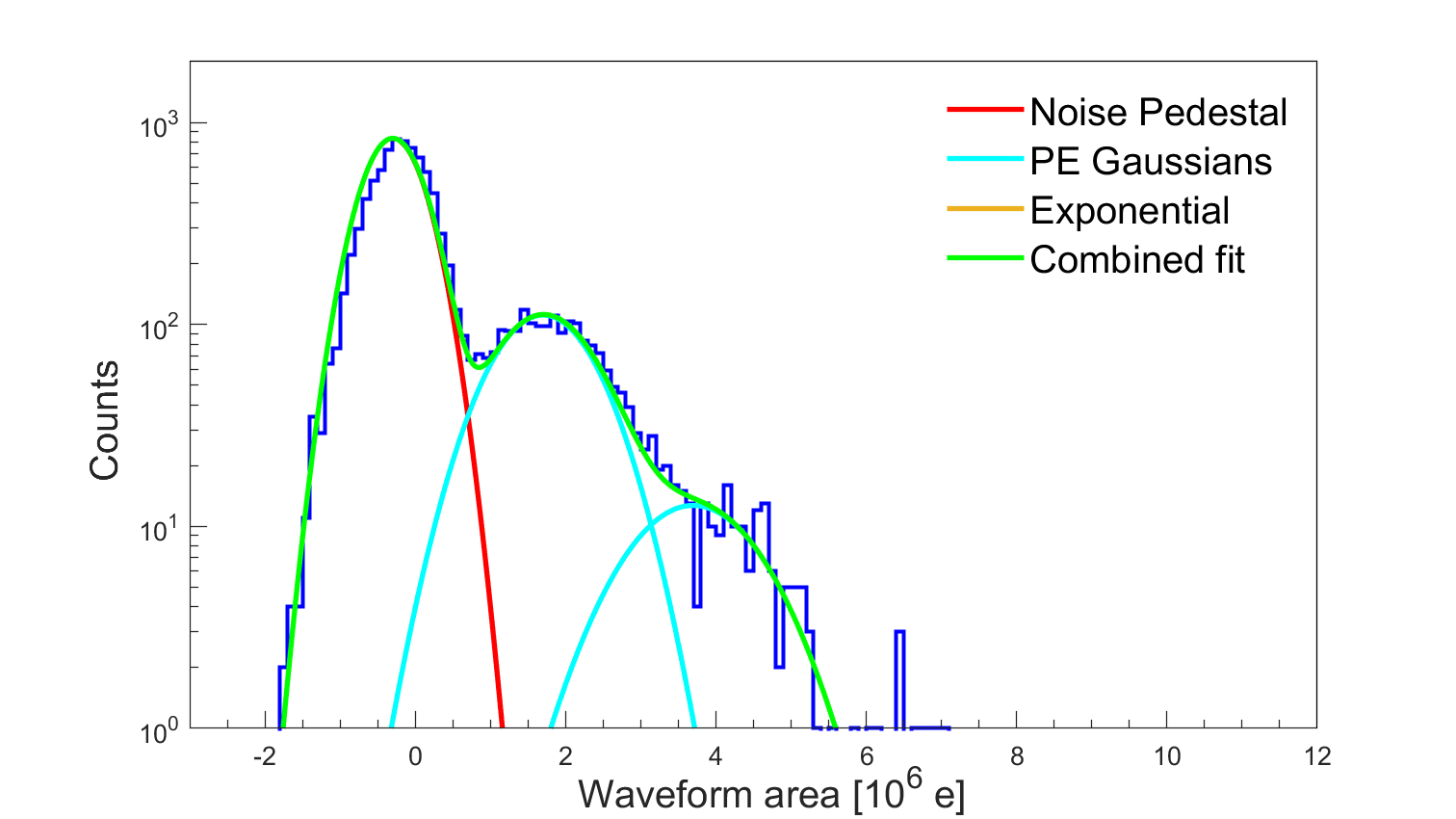}
    {\small\textbf{(f)}}\\[-0.5mm]
  \end{minipage}

  \vspace{2mm}
  \begin{minipage}{0.49\textwidth}
    \centering
    \includegraphics[width=\linewidth]{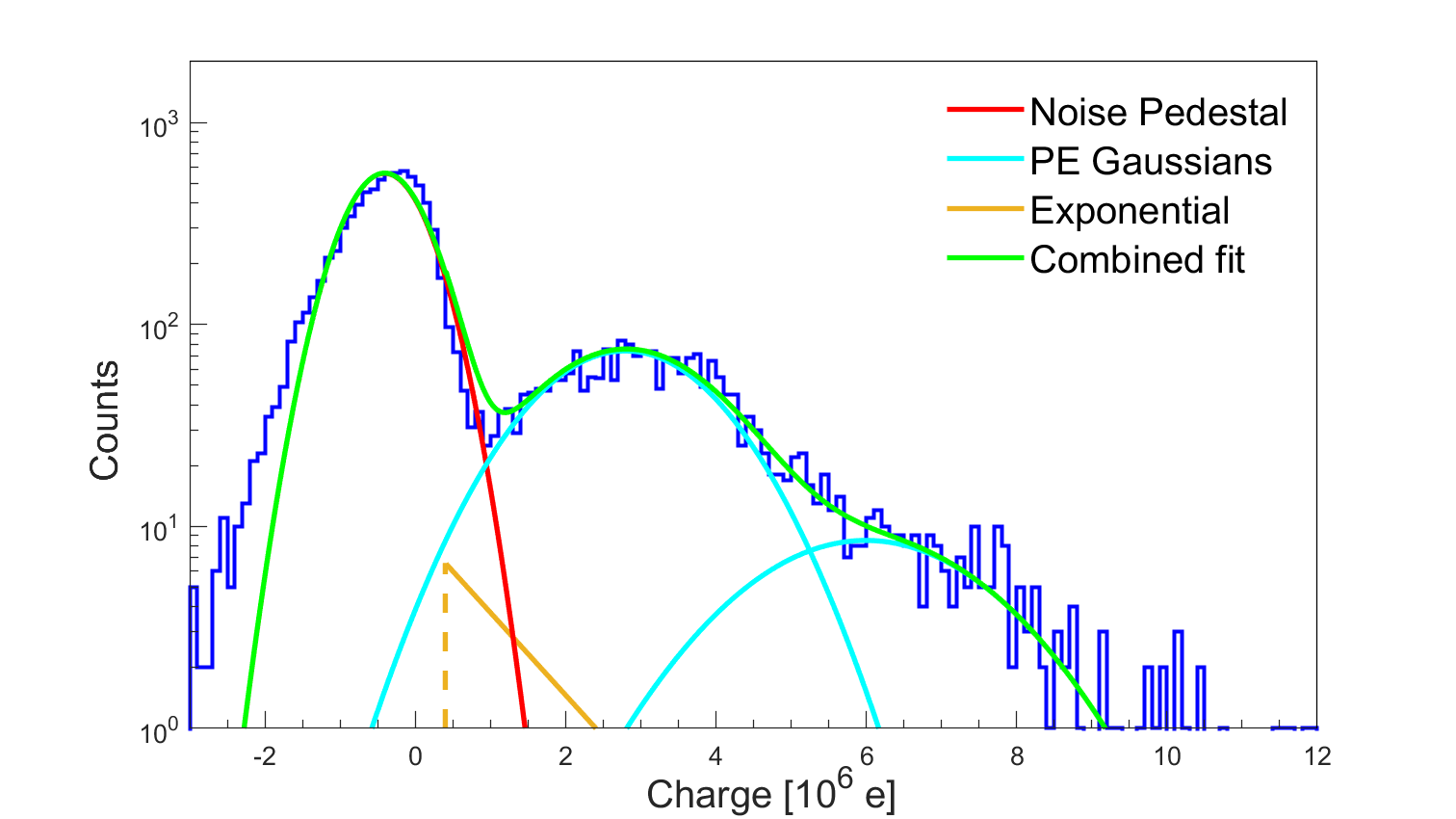}
    {\small\textbf{(c)}}\\[-0.5mm]
  \end{minipage}\hfill
  \begin{minipage}{0.49\textwidth}
    \centering
    \includegraphics[width=\linewidth]{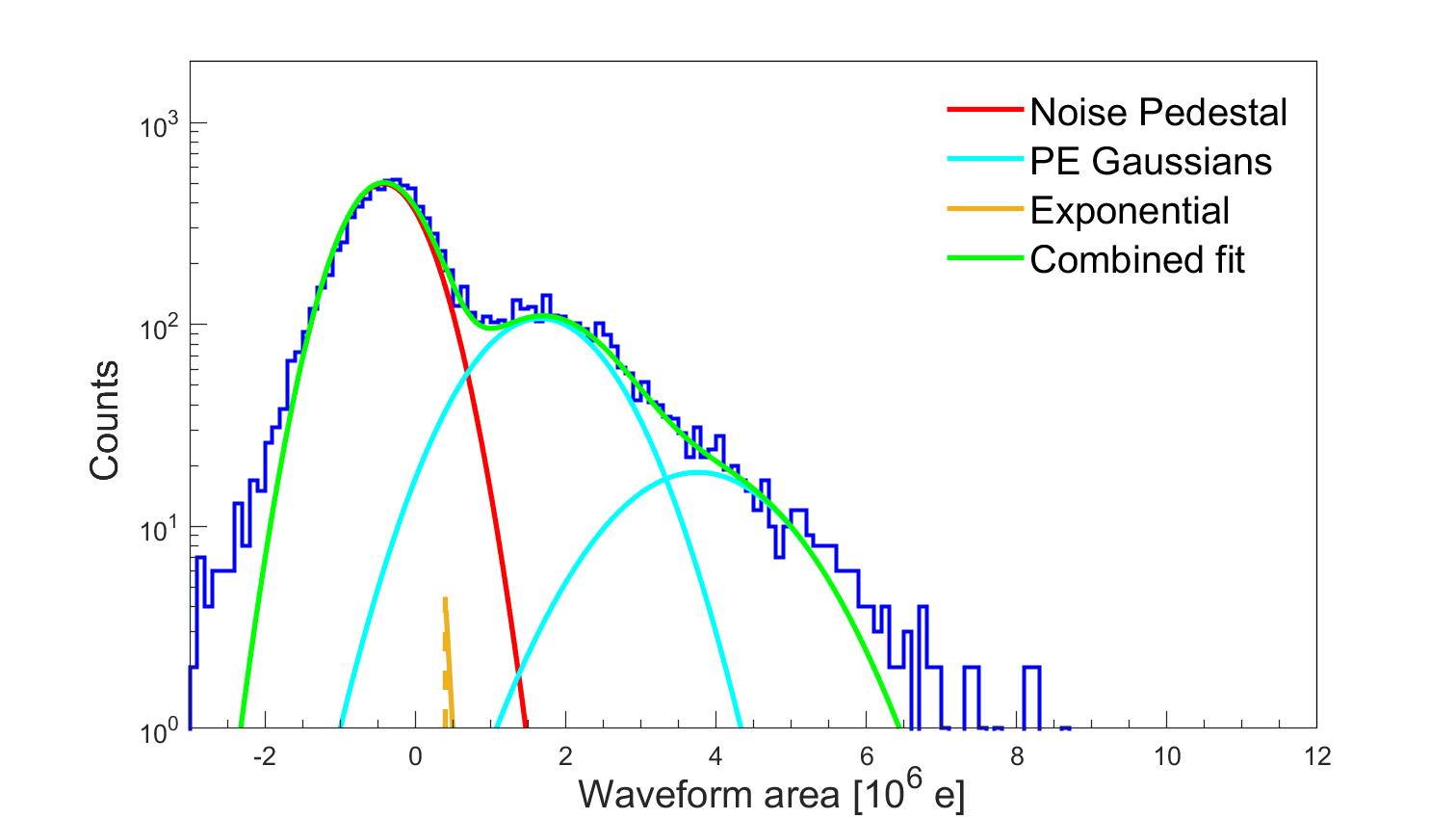}
    {\small\textbf{(g)}}\\[-0.5mm]
  \end{minipage}

  \vspace{2mm}
  \begin{minipage}{0.49\textwidth}
    \centering
    \includegraphics[width=\linewidth]{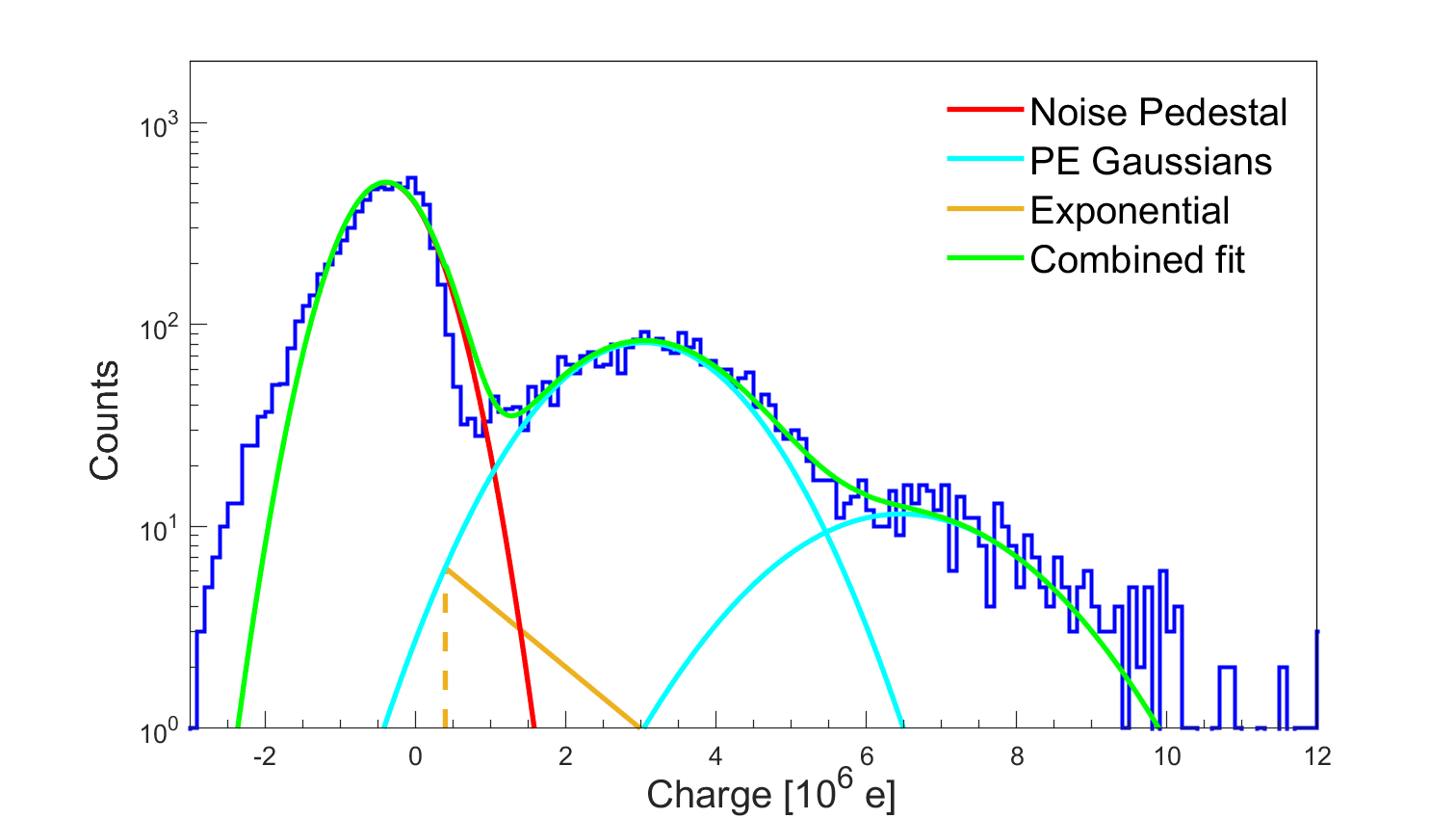}
    {\small\textbf{(d)}}\\[-0.5mm]
  \end{minipage}\hfill
  \begin{minipage}{0.49\textwidth}
    \centering
    \includegraphics[width=\linewidth]{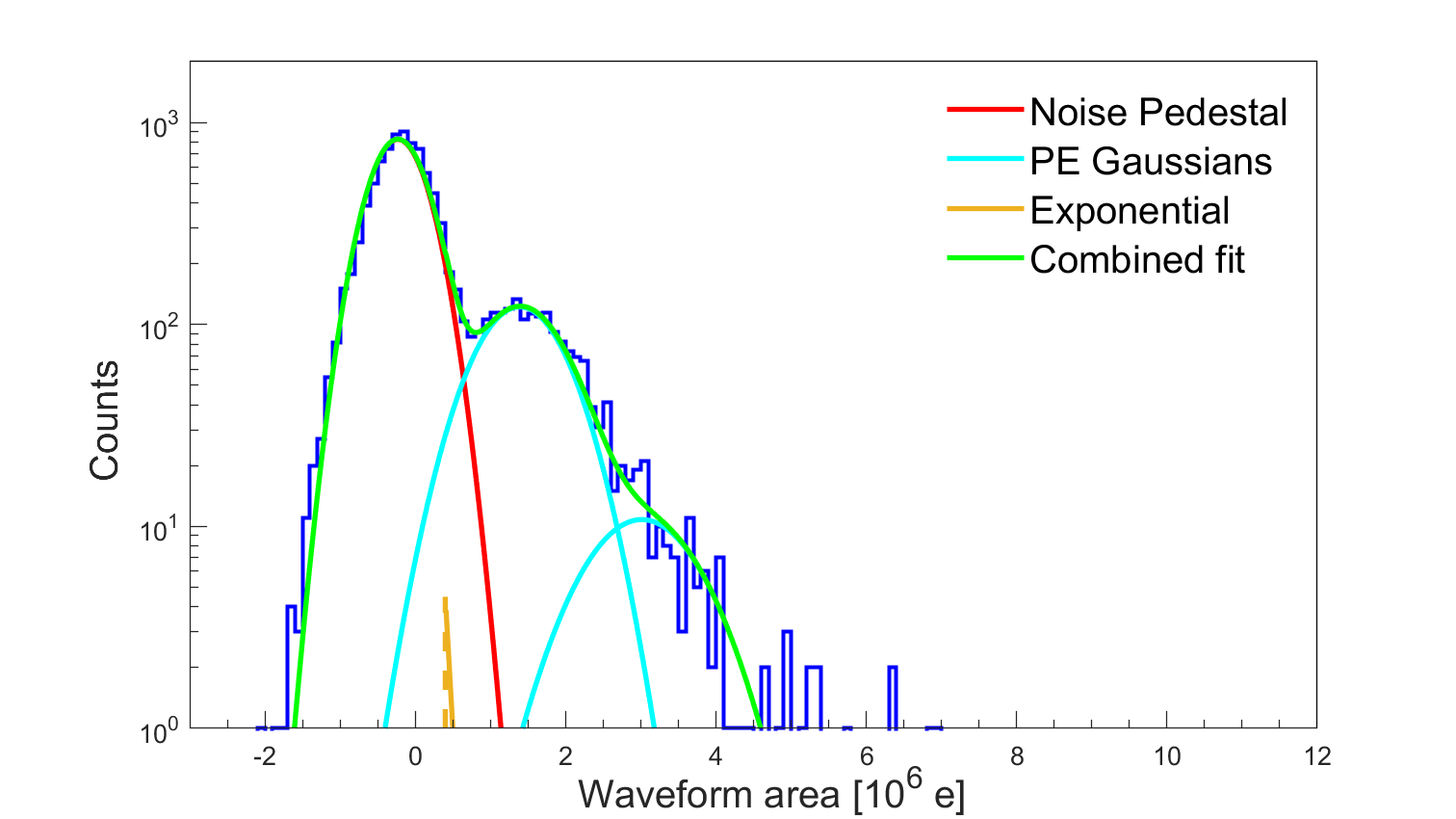}
    {\small\textbf{(h)}}\\[-0.5mm]
  \end{minipage}

  \caption{(a)-(d) The coaxial readout charge spectrum of an R12699 PMT's four channels. The gain values of the four channels are $3.40\times10^6$, $3.43\times10^6$, $3.20\times10^6$ and $3.43\times 10^6$ respectively. The SPE SNRs of the four channels are 6.42, 6.86, 6.09, and 6.13. (e)-(h) The optical readout waveform area spectrum. The mean values of SPE distributions are $2.86\times10^6$, $1.95\times10^6$, $2.09\times10^6$ and $1.63\times10^6$, while the corresponding SPE SNRs are 4.50, 5.00, 3.89, and 4.44.
  }
  \label{fig:coaxial_optical}
\end{figure}

According to Sec.~\ref{subsec:rematch}, the 1510~nm, 1470~nm, 1430~nm, and 1350~nm LDs form a four-channel multiplexed readout path through their re-matched CWDM channels at -100~$^{\circ}$C. During the test, the PMT was illuminated by an LED driven by periodic square-wave pulses generated by a waveform generator. The same trigger signal was used to synchronously trigger the oscilloscope, which recorded data from either the coaxial-cable readout or the optical readout. Following the same fitting procedure introduced in~\cite{MyR12699article}, identical fitting functions were applied to the charge and area spectra measured from the same PMT channel using the coaxial and optical readout methods. The comparison results are shown in Fig.~\ref{fig:coaxial_optical}.

From the spectra, the optical readout signals can be calibrated relative to the coaxial readout. The calibration factor is calculated as the mean value ratio of the fitted SPE distributions, obtained from the coaxial and optical readout spectra. Owing to differences in optical output power and coupling efficiency between the laser diodes and the CWDM channels, the calibration factors vary between channels. For the channels shown in Fig.~\ref{fig:coaxial_optical}~(a) and (b), the calibration factors are 1.188, 1.761, 1.530, and 2.105, respectively. The SPE SNR, defined as the SPE distribution mean value divided by the noise pedestal standard deviation, is measured to be 4.50, 5.00, 3.89, and 4.44 for the four multiplexed channels, with an average of 4.46.

\section{Summary}
\label{sec:sum}

In this work, a novel analog optical readout approach for low-temperature experiments is presented and experimentally validated, and the prototype achieves stable operation over a temperature range from 20~$^{\circ}$C down to -120~$^{\circ}$C. The developed prototype demonstrates a -3~dB bandwidth exceeding 150~MHz, a dynamic range larger than 500~mV (corresponding to approximately 80~times the SPE signal amplitude), and a cryogenic-region power consumption of 70~mW per channel.

Furthermore, by integrating optical transmission with optical wavelength division multiplexing, four-channel multiplexed readout through a single optical fiber is achieved at -100~$^{\circ}$C. A commissioning with an R12699 photomultiplier tube confirms the feasibility of the multiplexed analog optical readout, with the average SPE SNR larger than 4. Overall, the proposed multiplexed analog optical readout provides a viable solution for signal readout in low-temperature experiments.

\acknowledgments

This work was supported by the Ministry of Science and Technology of China (No.: 2023YFA1606203), Shanghai Pilot Program for Basic Research — Shanghai Jiao Tong University (No.: 21TQ1400218), Yangyang Development Fund, and National Natural Science Foundation of China (Grant No.: 12575197).

\bibliographystyle{JHEP}
\bibliography{biblio.bib}

\end{document}